\documentclass[10pt,twocolumn,letterpaper]{article}

\usepackage{cvpr}
\usepackage[dvipsnames]{xcolor}
\usepackage[toc]{appendix}

\usepackage{times}
\usepackage{epsfig}
\usepackage{graphicx}
\usepackage{amsmath}
\usepackage{amssymb}
\usepackage{float}
\usepackage{caption}
\usepackage{array,colortbl,multirow,multicol,booktabs,ctable}
\usepackage[percent]{overpic}
\usepackage{url}
\usepackage{enumitem}

\newcommand{\Vcent}[1]{\textcolor{orange}{#1}}
\newcommand{\Pcent}[1]{\textcolor{Goldenrod}{#1}}
\newcommand{\TenEighty}[1]{\textcolor{black}{#1}} 
\newcommand{\TtV}[1]{\textcolor{YellowGreen}{#1}} %
\newcommand{\TtX}[1]{\textcolor{NavyBlue}{#1}}
\newcommand{\TtXp}[1]{\textcolor{SkyBlue}{#1}}
\newcommand{\Mix}[2]{\textit{\TenEighty{#1}\TenEighty{#2}}*}

\usepackage[breaklinks=true,bookmarks=false]{hyperref}

\cvprfinalcopy 

\usepackage{stfloats}
\begin{document}

\title{ABC: A Big CAD Model Dataset For Geometric Deep Learning}

\author{
\parbox{0.28\textwidth}{
\centering
Sebastian Koch\\
TU Berlin\\
{\tt\small s.koch@tu-berlin.de}
}
\and
\parbox{0.28\textwidth}{
\centering
Albert Matveev\\
Skoltech, IITP\\
{\tt\small albert.matveev@skoltech.ru}
}
\and
\parbox{0.28\textwidth}{
\centering
Zhongshi Jiang\\
New York University\\
{\tt\small jiangzs@nyu.edu}
}
\and
\parbox{0.28\textwidth}{
\centering
Francis Williams\\
New York University\\
{\tt\small francis.williams@nyu.edu}
}
\and
\parbox{0.28\textwidth}{
\centering
Alexey Artemov\\
Skoltech\\
{\tt\small a.artemov@skoltech.ru}
}
\and
\parbox{0.28\textwidth}{
\centering
Evgeny Burnaev\\
Skoltech\\
{\tt\small e.burnaev@skoltech.ru}
}
\and
\parbox{0.28\textwidth}{
\centering
Marc Alexa\\
TU Berlin\\
{\tt\small marc.alexa@tu-berlin.de}
}
\and
\parbox{0.28\textwidth}{
\centering
Denis Zorin\\
NYU, Skoltech\\
{\tt\small dzorin@cs.nyu.edu}
}
\and
\parbox{0.28\textwidth}{
\centering
Daniele Panozzo\\
New York University\\
{\tt\small panozzo@nyu.edu}
}
}

\maketitle

\newcommand{\todo}[1]{{\color{red} #1}}

\newcommand{\ra}[1]{\renewcommand{\arraystretch}{#1}}
\newcommand{\cm}[0]{\checkmark}

\newcommand{\mv}[1]{\mathbf{#1}}

\ra{1.04}
\setlength{\tabcolsep}{5.5pt}

\setlist[description]{font=\normalfont\itshape\space}

\makeatletter
\renewcommand{\paragraph}{%
  \@startsection{paragraph}{4}%
  {\z@}{1.2ex \@plus 1ex \@minus .2ex}{-1em}%
  {\normalfont\normalsize\bfseries}%
}
\makeatother




\begin{abstract}
\vspace{-2mm}

We introduce ABC-Dataset, a collection of one million Computer-Aided Design (CAD) models for research of  geometric deep learning methods and applications. Each model is a collection of explicitly parametrized curves and surfaces,
providing ground truth for differential quantities, patch segmentation, geometric feature detection, and shape reconstruction. Sampling the parametric descriptions of surfaces and curves allows generating data in different formats and resolutions, enabling fair comparisons for a wide range of geometric learning algorithms. 
As a use case for our dataset, we perform a large-scale benchmark for estimation of surface normals, comparing existing data driven methods and evaluating their performance against both the ground truth and traditional normal estimation methods.

\end{abstract}

\begin{figure*}[b]
\vspace{-5mm}
\includegraphics[width=1.0\linewidth]{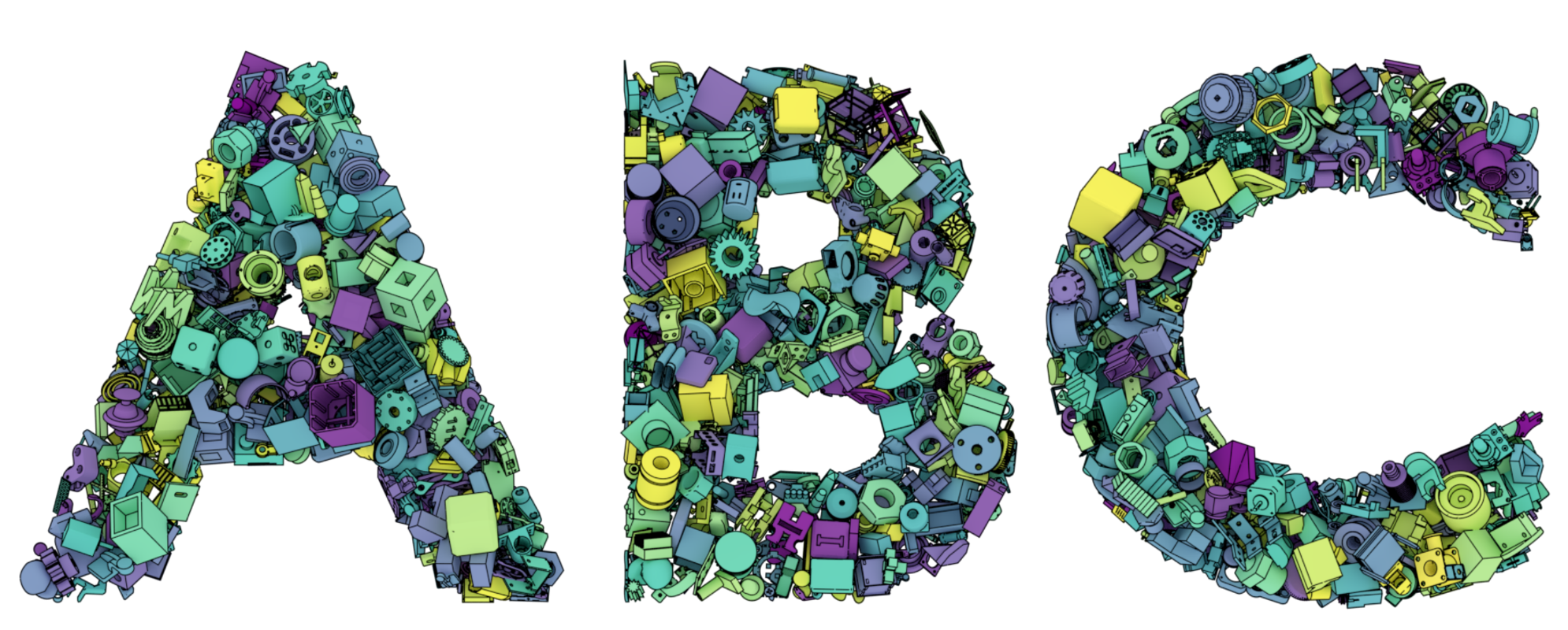}
\caption{Random CAD models from the ABC-Dataset: \url{https://deep-geometry.github.io/abc-dataset}}\label{fig:abc}
\vspace{-5mm}
\end{figure*}

\section{Introduction}

The combination of large data collections and deep learning algorithms is transforming many areas of computer science. Large data collections are an essential part of this transformation. Creating these collections for many types of data  (image, video, and audio) has been boosted  by the ubiquity of acquisition devices and mass sharing of these types of data on social media. In all these cases, the data representation is based on regular discretization in space and time providing structured and uniform input for deep learning algorithms.

The situation is different for three-dimensional geometric models. Acquiring or creating high-quality models of this type is still difficult, despite growing availability of 3D sensors, and improvements in 3D design tools.  Inherent irregularity of the surface data, and still-significant level of skill needed for creating high-quality 3D shapes contributes to the limited availability of geometric datasets. Irregularity of geometric data is reflected in commonly used geometric formats, which differ in fundamental ways from formats for images, video, and audio.  Existing datasets often lack reliable ground truth annotations. We further discuss currently available geometric datasets in Section~\ref{ssec:datasets}.

Common shape analysis and geometry processing tasks that can benefit from \emph{geometric deep learning} include estimation of differential surface properties (Section \ref{sec:evaluation}), feature detection, and shape reconstruction. Ground truth for some of these tasks is hard to generate, as marking features by hand is a laborious task and differential properties can only be approximated for sampled surfaces.

In this work, we make the following contributions:

\paragraph{Dataset.} We introduce a dataset for geometric deep learning consisting of over 1 million individual (and high quality) geometric models, each defined by parametric surfaces and associated with accurate ground truth information on the decomposition into patches, sharp feature annotations, and analytic differential properties. We gather the models through a publicly available interface hosted by Onshape \cite{Onshape}. 
Similar to vector graphics for images, this representation allows resampling the surface data at arbitrary resolutions, with or without connectivity information (i.e. into a point cloud or a mesh).  

\paragraph{Benchmark.} We demonstrate the use of the dataset by building a benchmark for the estimation of surface normals. This benchmark is targeting methods that compute normals (1) locally on small regions of a surface, and (2) globally over the entire surface simultaneously. We have chosen this problem for two reasons: most existing geometric deep learning methods were tested on this task, and very precise ground truth can be readily obtained for shapes in our data set. We run the benchmark on 7 existing deep learning algorithms, studying how they scale as the dataset size increases, and comparing them with 5 traditional geometry processing algorithms to establish an objective baseline for future algorithms. Results are presented in Section~\ref{sec:evaluation}.

\paragraph{Processing Pipeline.} We develop an open-source geometry processing pipeline that processes the CAD models to directly feed deep learning algorithms. Details are provided in Section~\ref{sec:dataset}.
We will continually update the dataset and benchmark as more models are added to the public collection of models by Onshape. Our contribution adds a new resource for the development of geometric deep learning, targeting applications focusing on human-created, mechanical shapes. It will allow researchers to compare against existing techniques on a large and realistic dataset of man-made objects.

\section{Related Work}

We review existing datasets for data-driven processing of geometrical data, and then review both data-driven and analytical approaches to estimate differential qualities on smooth surfaces.

\paragraph{3D Deep Learning Datasets.}
\label{ssec:datasets}
The community has seen a growth in the availability of 3D models and datasets. Segmentation and classification algorithms have benefited greatly from the most prominent ones \cite{shapenet,modelnet,Thingi10K,psb}. Further datasets are available for large scenes \cite{sunrgbd, scannet}, mesh registration \cite{dfaust} and 2D/3D alignment \cite{chair3d}. The dataset proposed in this paper has the unique property of containing the analytic representation of surfaces and curves, which is ideal for a quantitative evaluation of data-driven methods. Table~\ref{tab:datasets} gives an overview of the most comparable datasets and their characteristics and capabilities.

\begin{table}
\begin{center}
\begin{tabular}{llllllr}
Dataset & \#Models & \rotatebox[origin=c]{90}{CAD Files} & \rotatebox[origin=c]{90}{Curves} & \rotatebox[origin=c]{90}{Patches} & \rotatebox[origin=c]{90}{Semantics} & \rotatebox[origin=c]{90}{Categories}\\
\midrule
ABC                     & 1,000,000+   & \cm   & \cm   & \cm   & --    & --\\
ShapeNet* \cite{shapenet}  & 3,000,000+   & --    & --    & --     & \cm   & \cm \\
ShapeNetCore            & 51,300       & --    & --    & --     & \cm   & \cm \\
ShapeNetSem             & 12,000       & --    & --    & --     & \cm   & \cm \\
ModelNet  \cite{modelnet}  & 151,128      & --    & --    & --     & \cm   & \cm \\
Thingi10K  \cite{Thingi10K}  & 10,000       & --    & --    & --    & --    & \cm\\
PrincetonSB\cite{psb}             & 6670         & --    & --    & --    & --    & \cm\\
NIST \cite{nistcad}  & $\leq30$     & \cm   & \cm   & \cm   & --    & --\\
\midrule
\end{tabular}
\end{center}
\vspace{-2mm}
\caption{Overview of existing datasets and their capabilities. *: The full ShapeNet dataset is not yet publicly available, only the subsets ShapeNetCore and ShapeNetSem.}
\label{tab:datasets}
\end{table}

\paragraph{Point Cloud Networks.}

Neural networks for point clouds are particularly popular, as they make minimal assumptions on input data.  One of the earliest examples is PointNet \cite{pointnet} and its extension PointNet++ \cite{pointnetpp}, which ensure that the network output is invariant with respect to point permutations.
PCPNet \cite{pcpnet} is a variation of PointNet tailored for estimating local shape properties: it extracts local patches from the point cloud, and estimates local shape properties at the central points of these patches. 
PointCNN \cite{pointcnn} explores the idea of learning a transformation from the initial point cloud, which provides the weights for input points and associated features, and produces a permutation of the points into a latent order.

Point Convolutional Neural Networks by Extension Operators \cite{pcnnExt} is a fundamentally different way to process point clouds through mapping point cloud functions to volumetric functions and vice versa through extension and restriction operators. A similar volumetric approach has been proposed in Pointwise Convolutional Neural Networks \cite{pwcnn} for learning pointwise features from point clouds.

PointNet-based techniques, however, do not attempt to use local neighborhood information explicitly.  Dynamic Graph CNNs \cite{dgcnn} uses an operation called EdgeConv, which exploits local geometric structures of the point set by generating a local neighborhood graph based on proximity in the feature space and applying convolution-like operation on the edges of this graph. 
In a similar fashion, FeaStNet \cite{feastnet} proposes a convolution operator for general graphs, which applies filter kernels in a data-driven manner to the local irregular neighborhoods.

\paragraph{Networks on Graphs and Manifolds.}
Neural networks for graphs have been introduced in  \cite{scarselli2009graph}, and extended in \cite{li2015gated, sukhbaatar2016learning}. A wide class of convolutional neural networks with spectral filters on graphs was introduced in \cite{bruna2013spectral} and developed further in \cite{henaff2015deep, defferrard2016convolutional, kipf2016semi}. Graph convolutional neural networks were applied to non-rigid shape analysis in \cite{boscaini2015learning,yi2017syncspeccnn}. 
Surface Networks \cite{surface_networks} further proposes the use of the differential geometry operators to extend GNNs to exploit properties of the surface. 
For a number of problems where quantities of interest are localized, spatial filters are more suitable than spectral filters. A special CNN model for meshes which uses intrinsic patch representations was presented in \cite{masci2015geodesic}, and further generalized in \cite{boscaini2016learning, monti2017geometric}. In contrast to these intrinsic models, Euclidean models \cite{wu20153d, wei2016dense} need to learn the underlying invariance of the surface embedding, hence they have higher sample complexity.
More recently, \cite{maron2017convolutional} presented a surface CNN based on the canonical representation of planar flat-torus. 
We refer to \cite{bronstein2017geometric} for an extensive overview of geometric deep learning methods.

\paragraph{Analytic Normal Estimation Approaches.}

The simplest methods are based on fitting tangent planes. For a point set, these methods estimate normals in the points as directions of smallest co-variance (i.e., the tangent plane is the total least squares fit to the points in the neighborhood). For a triangle mesh, normals of the triangles adjacent to a vertex can be used to compute a vertex normal as a weighted average. We consider uniform, area, and angle weighting~\cite{normal_comparison}.

One can also fit higher order surfaces to the discrete surface data. These methods first estimate a tangent plane, and then fit a polynomial over the tangent plane that interpolates the point for which we want to estimate the normal (a so-called osculating jet~\cite{osculatingjets}). A more accurate normal can then be recomputed from the tangents of the polynomial surface approximation at the point. The difference for triangle meshes and point sets is only in collecting the samples in the neighborhood. In addition, one can use robust statistics to weight the points in the neighborhood~\cite{discrete_surf}.

For the weighted triangle normals we use the implementation of libigl~\cite{libigl}, for computation of co-variance normals and osculating jets we use the functions in CGAL~\cite{cgal:pc-eldp-18b,cgal:ass-psp-18b}. For the robust estimation the authors have provided source code.
There are many other techniques for estimating surface normals (e.g.~\cite{mura}), however we only focus on a selected subset in the current work.

One important generalization is the detection of sharp edges and corners, where more than one normal can be assigned to a point~\cite{Ohtake:2003:MPU,fastrobustsharp}. However, as we only train the machine learning methods to report a single normal per point, we leave an analysis of these extensions for future work. 

\section{Dataset}
\label{sec:dataset}
We identify six crucial properties that are desirable for an "ideal" dataset for geometric deep learning: (1) large size: since deep networks require large amounts of data, we want to have enough models to find statistically significant patterns; (2) ground truth signals: in this way, we can quantitatively evaluate the performance of learning on different tasks; (3) parametric representation: so that we can resample the signal at the desired resolution without introducing errors; (4) expandable: it should be easy to make the collection grow over time, to keep the dataset challenging as progress in learning algorithms is made; (5) variation: containing a good sampling of diverse shapes in different categories; (6) balanced: each type of objects should have a sufficient number of samples. 

Since existing datasets are composed of acquired or synthesized point clouds or meshes, they do not satisfy property 2 or 3. We thus propose a new dataset of CAD models, which is complementary: it satisfies properties 1-4; it is restricted to a specific type of models (property 6), but has a considerable variation inside this class.  While restriction to CAD models can be viewed as a downside, it strikes a good balance between having a sufficient number of similar samples and diversity of represented shapes, in addition to having high-quality ground truth for a number of quantities. 

\begin{figure*}[t]
\begin{center}
\includegraphics[width=1.0\linewidth]{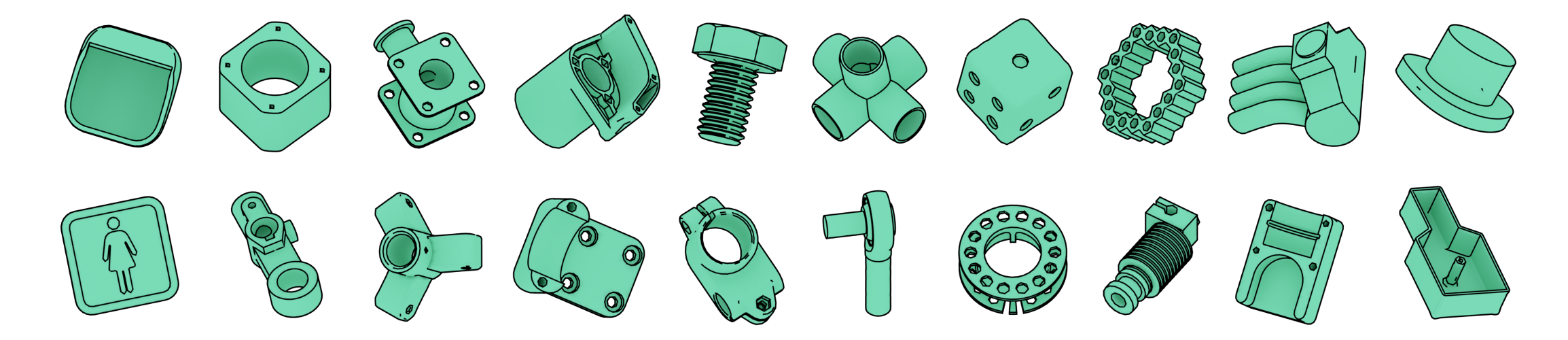}

\end{center}
\vspace{-3mm}
   \caption{Random examples from the dataset. Most models are mechanical parts with sharp edges and well defined surfaces.}
\label{fig:models}

\end{figure*}

\paragraph{Acquisition.} Onshape has a massive online collection of CAD models, which are freely usable for research purposes. By collecting them over a period of 4 months we obtained a collection of over 1 million models (see Figure~\ref{fig:models}). 

\paragraph{Ground Truth Signals and Vector Representation.} The data is encoded in a vectorial format that enables to resample it at arbitrary resolution and to compute analytically a large set of signals of interest (Section \ref{sec:applications}), which can be used as a ground truth.

\paragraph{Challenge.} However, the data representation is not suitable for most learning methods, and the conversion of CAD models to discrete formats is a difficult task. This paper presents a robust pipeline to process, and use CAD models as an ideal data source for geometry processing algorithms.

\subsection{CAD Models and Boundary Representations}
In the following, we will use the term CAD model to refer to a geometric shape defined by a boundary representation (B-Rep). The boundary representation describes models by their topology (faces, edges, and vertices) as well as their geometry (surfaces, curves, and points).
The topology description is organized in a hierarchical way:
\emph{Solids} are bound by a set of oriented faces which are called shells;
\emph{Faces} are bound by wires which are ordered lists of edges (the order defines the face orientation);
\emph{Edges} are the oriented connections between 2 vertices;
\emph{Vertices} are the basic entities, corresponding to points in space.

Each of these entities has a geometric description, which allows us to embed the model in 3D space. In our dataset, each surface can represent a plane, cone, cylinder, sphere, torus, surface of revolution or extrusion or NURBS patch \cite{Farin:2001}. Similarly, curves can be lines, circles, ellipses, parabolas, hyperbolas or NURBS curves \cite{Farin:2001}. Each vertex has coordinates in 3D space. An additional complexity is added by trimmed patches (see \cite{Farin:2001} for a full description).

\subsection{Processing Pipeline}
\label{ssec:geoproc}
Assembling a dataset of such a large size is a time-consuming task where many design decisions have to be made beforehand: as an example, it requires around 2 CPU years to extract triangle meshes for 1 million CAD models.

To encourage active community participation and easy adoption, we use accessible, well-supported open-source tools, instead of relying on commercial CAD software. It allows the community to use and expand our dataset in the future. Our pipeline is designed to run in parallel on large computing clusters. 

The Onshape public collection is not curated. It contains all the public models created by their users, without any additional filtering. Despite the fact that all models have been manually created, there is a small percentage of imperfect models with broken boundaries, self-intersecting faces or edges, as well as duplicate vertices. In addition to that, there are many duplicate models, and especially models that are just made of single primitives such as a plane, box or cylinder, probably created by novice users that were learning how to use Onshape.

Given the massive size of the dataset, we developed a set of geometric and topological criteria to filter low quality of defective models, which we describe in the supplementary material, and we leave a semantic, crowd-sourced filtering and annotation as a direction for future work.

\begin{enumerate}[label=,start=1,leftmargin=*]
\item \textbf{Step 1: B-Rep Loading and Translation.} The STEP files \cite{step} we obtain from Onshape contain the boundary representation of the CAD model, which we load and query using the open-source software Open Cascade \cite{Opencascade}. 
The translation process generates for each surface patch and curve segment an explicit parameterization, which can be sampled at arbitrary resolution.
\item \textbf{Step 2: Meshing/Discretization.} The parameterizations of the patches are then sampled and triangulated using the open-source software Gmsh \cite{gmsh}. We offer an option here to select between uniform (respecting a maximal edge length) and curvature adaptive sampling.
\item \textbf{Step 3: Topology Tree Traversal/Data Extraction.}
The sampling of B-Rep allows to track correspondences between the continuous and discrete representations. As they can be differentiated at arbitrary locations, it is possible to calculate ground truth differential quantities for all samples of the discrete model. 
Another advantage is the explicit topology representation in B-Rep models, which can be transferred to the mesh in the form of labels. These labels define for each triangle of the discrete mesh to which surface patch it belongs. The same applies also for the curves, we can label edges in the discrete mesh as sharp feature edges. 
While CAD kernels provide this information, it is difficult to extract it in a format suitable for learning tasks. Our pipeline exports this information in yaml files with a simple structure (see supplementary material).
\item \textbf{Step 4: Post-processing.}
We provide tools to filter our meshes depending on quality or number of patches, to compute mesh statistics, and to resample the generated surfaces to match a desired number of vertices \cite{Garland:1997}. 
\end{enumerate}

\subsection{Analysis and Statistics}

\begin{figure}[t]
\begin{center}
\includegraphics[width=1.0\linewidth]{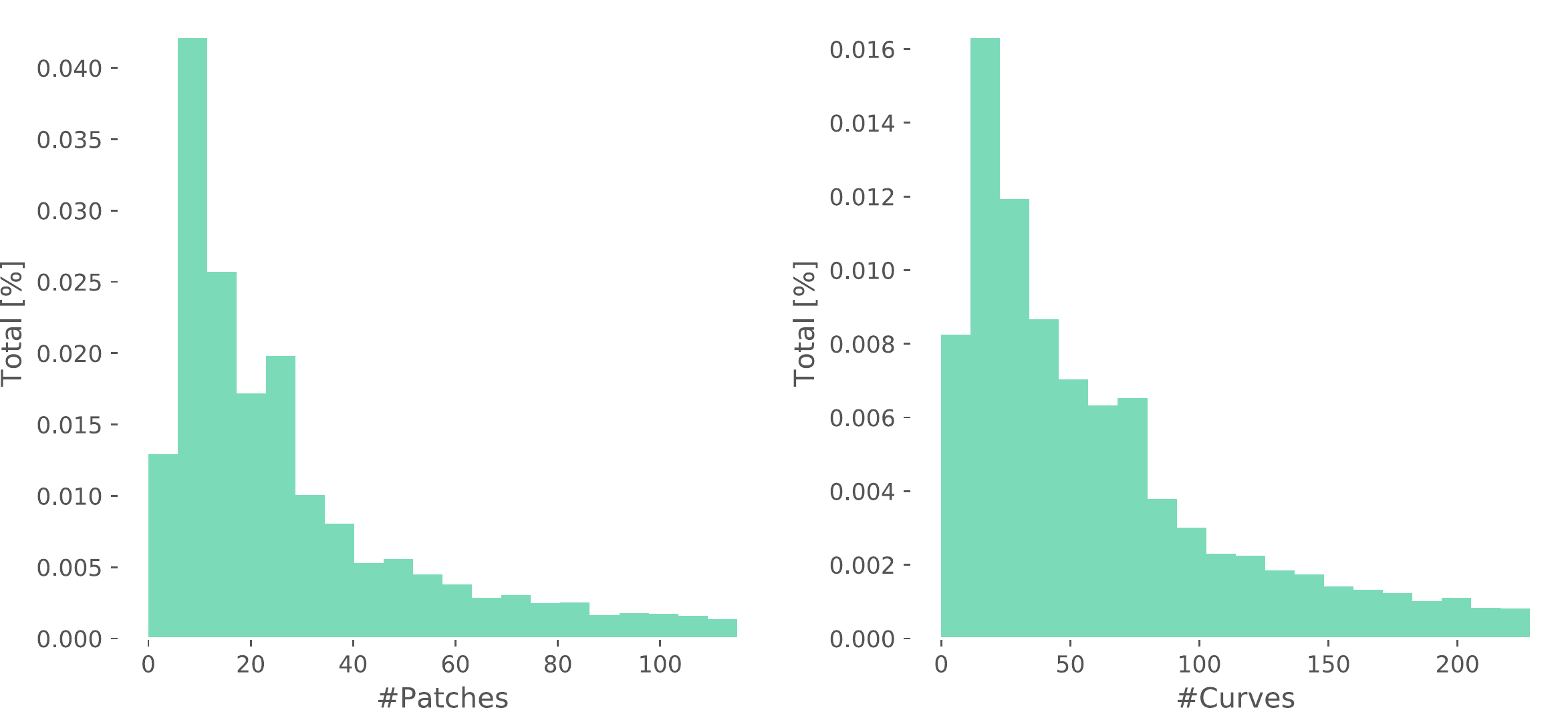}
\end{center}
\vspace{-3mm}
   \caption{Histograms over the number of patches and curves per CAD model. This shows that there are many simpler models which consist of less than 30 patches/100 curves as well as more complex ones. Both histograms are truncated at the right side.}
\label{fig:complexity}
\end{figure}

We show an overview of the models in the dataset in Figures \ref{fig:abc} and \ref{fig:models}. 
In Figure~\ref{fig:typedistribution} and \ref{fig:complexity} we show the distribution of surface and edge types and the histogram of patch and edge numbers, to give an impression of the complexity and variety of the dataset.
Updated statistics about the growing dataset are available on our dataset website \cite{abc}.

\begin{figure}[t]
\begin{center}
\includegraphics[width=1.0\linewidth]{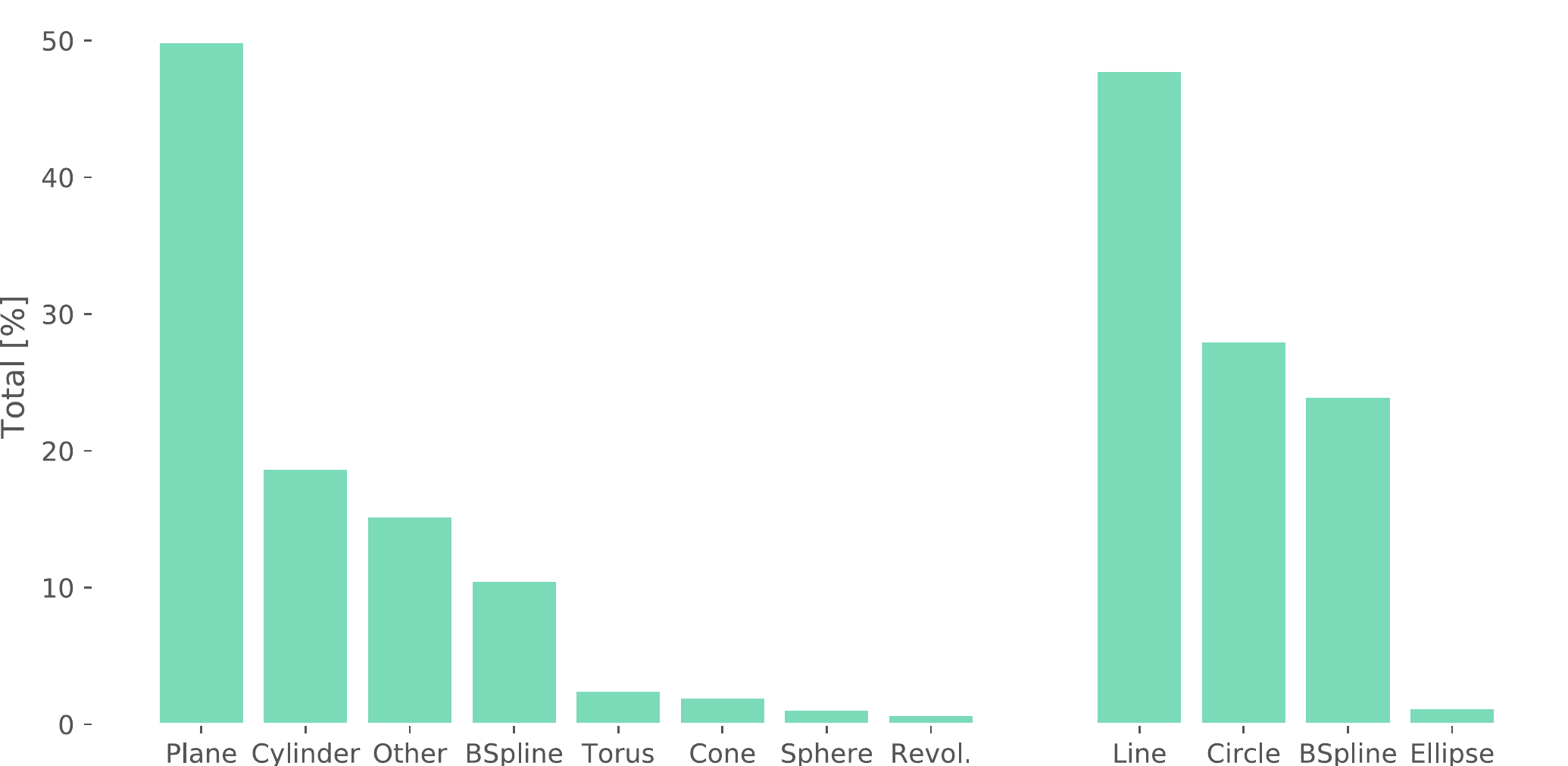}
\end{center}
\vspace{-2mm}
   \caption{Each model in our dataset is composed of multiple patches and feature curves. The two images show the distribution of types of patches (left) and curves (right) over the current dataset ($\approx 1$M models). }
\label{fig:typedistribution}
\end{figure}

\section{Supported Applications}
\label{sec:applications}

We briefly overview a set of applications that may benefit from our dataset, that can be used as either training data or as a benchmark.

\paragraph{Patch Decomposition.} Each object in our collection is naturally divided into surface regions, separated by feature lines. The decomposition is defined by the author when a shape is constructed, and is likely to be semantically meaningful. It is also constrained by a strong geometric criteria: each region should be representable by a (possibly trimmed) NURBS patch. 

\paragraph{Surface Vectorization.}
The B-rep of a CAD models is the counterpart of a vector representation for images, that can be resampled at any desired resolution. The conversion of a surface triangle mesh into a B-rep is an interesting and challenging research direction, for which data driven methods are still at their infancy  \cite{Shapiro:1991,Sharma:2018,Du2018InverseCSG}.

\paragraph{Estimation of Differential Quantities.}
Our models have ground truth normals and curvature values making them an ideal, objective benchmark for evaluating algorithms to predict these quantities on point clouds or triangle meshes of artificial origin.

\paragraph{Sharp Feature Detection.}
Sharp features are explicitly encoded in the topological description of our models, and it is thus possible to obtain ground truth data for predicting sharp features on point clouds \cite{Weber:2010} and meshes.

\paragraph{Shape Reconstruction.}
Since the ground truth geometry is known for B-rep models, they can be used to simulate a scanning setup and quantitatively evaluate the reconstruction errors of both reconstruction \cite{berger2013benchmarkSR, sunnoise} and point cloud upsampling \cite{yu2018pu} techniques.

\paragraph{Image Based Learning Tasks.}
Together with the dataset, we provide a rendering module (based on Blender \cite{Blender}) to generate image datasets. It supports rendering of models posed in physically static orientations on a flat plane, different lighting situations, materials, camera placements (half-dome, random) as well as different rendering modes (depth, color, contour). Note that all these images can be considered ground truth, since there is no geometric approximation error.

\paragraph{Robustness of Geometry Processing Algorithms.}
Besides data-driven tasks, the dataset can also be employed for evaluating the robustness of geometry processing algorithms. Even a simple task like normal estimation is problematic on such a large dataset. Most of the methods we evaluated in Section~ \ref{sec:evaluation} fail (i.e. produce invalid normals with length zero or which contain NANs) on at least one model: our dataset is ideal for studying and solving these challenging robustness issues.

\section{Normal Estimation Benchmarks}

We now introduce a set of large scale benchmarks to evaluate algorithms to compute surface normals, exploiting the availability of ground truth normals on the B-rep models. To the best of our knowledge, this is the first large scale study of this kind, and the insights that it will provide will be useful for the development of both data-driven and analytic methods.

\begin{figure*}[t]
\begin{center}
\includegraphics[width=1.0\linewidth]{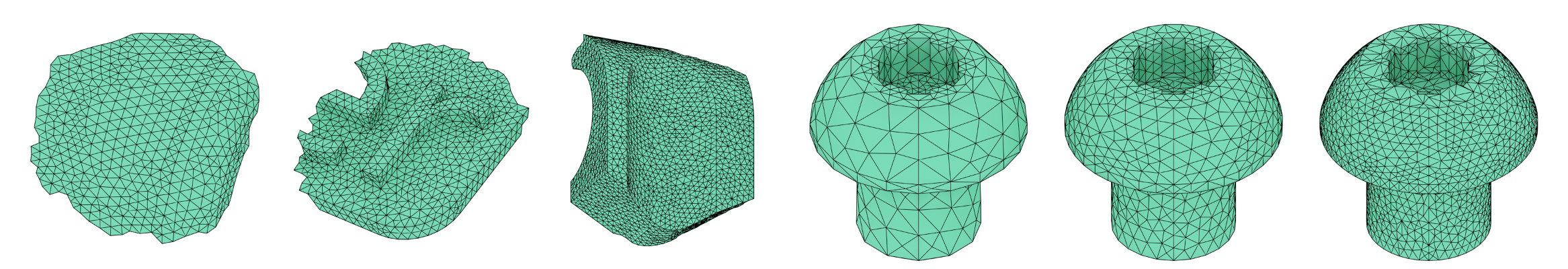}
\end{center}
\vspace{-3mm}
   \caption{Samples from the different categories in our normal estimation benchmark. From left to right: local patches of growing size and complexity (512, 1024, 2048 vertices), and full models at different densities (512, 1024, 2048 vertices).}
\label{fig:examples}
\end{figure*}

\paragraph{Construction.} To fairly compare a large variety of both data-driven and analytic methods, some targeting local estimation and some one-shot prediction of all normals of a 3D model, we build a series of datasets by randomly sampling points on our meshed B-reps, and growing patches of different sizes, ranging from 512 vertices to the entire model (Figure~\ref{fig:examples}). For each patch size, we generate 4 benchmarks with an increasing number of patches, from 10k to 250k, to study how the data-driven algorithms behave when they are trained on a larger input set.

\paragraph{Split.} All benchmark datasets are randomly split into training and test set with a distribution of $80\%$ training data and $20\%$ test data. The split will be provided to make the results reproducible.

\subsection{Evaluation}
\label{sec:evaluation}

\paragraph{Algorithms.}
We select 12 representative algorithms from the literature, and 5 of them are traditional ones, Robust Statistical Estimation on Discrete Surfaces (\textbf{RoSt}) \cite{discrete_surf} operating on point clouds (\textbf{PC}), and meshes (\textbf{M}), Osculating Jets (\textbf{Jets}) \cite{osculatingjets}, also on point clouds and meshes, and Uniform weighting of adjacent face normals (\textbf{Uniform}) \cite{normal_comparison}.
Notice that we replace K-ring neighborhoods with K-nearest neighbors for RoSt and Jets to support point cloud input.
Also, 7 machine learning methods are selected, 
including PointNet++ (\textbf{PN++}) \cite{pointnetpp}, 
Dynamic Graph CNN (\textbf{DGCNN}) \cite{dgcnn}, 
Pointwise CNN (\textbf{PwCNN}) \cite{pwcnn}, 
PointCNN (\textbf{PCNN}) \cite{pointcnn}, 
Laplacian Surface Network (\textbf{Laplace}) \cite{surface_networks}, 
PCP-Net (\textbf{PCPN}) \cite{pcpnet} 
and Point Convolutional Neural Networks by Extension Operators (\textbf{ExtOp}) \cite{pcnnExt}.
Of these methods, Laplace operates on triangle mesh input and the rest on point cloud input. Most of their output is one normal per vertex, except for PCPN and ExtOp the output is one normal for the center of the patch.
We provide a detailed explanation of the (hyper-)parameters and modifications we did for each method in the supplementary material.  For the statistics, we used only the valid normals reported by each method, and we filtered out all the degenerate ones.

\paragraph{Protocol.}
We compare the methods above on the benchmarks, using the following protocol: (1) for each method, we obtained the original implementation from the authors (or a reimplementation in popular libraries); (2) we used the recommended or default values for all (hyper-)parameters of the learning approaches (if these were not provided, we fine-tuned them on a case by case basis); (3) if the implementation does not directly support normal estimation, we modified their code following the corresponding description in the original paper (4) we used the same loss function $1 - \left(\mathbf{n}^\mathsf{T} \mathbf{\hat{n}}\right)^2$, with $\mathbf{n}$ as the estimated normal and $\mathbf{\hat{n}}$ as the ground truth normal, for all methods. Note that this loss function does not penalize normals that are inverted (flipped by $180^\circ$), which is an orthogonal problem usually fixed as in a postprocessing step \cite{hoppe1992surface}.

\paragraph{Results.}

A listing of the statistical results for all methods is given in Table~\ref{tab:results}.
Our experiments show that neural networks for normal estimation are stable across several runs; standard deviation of the losses is of the order of $10^{-3}$. For the 10k dataset, most of the networks converge within 24 hours on a NVIDIA GeForce GTX 1080 Ti GPU. We capped the training time of all methods to 72 hours.

\paragraph{Comparison of Data-Driven Methods.}

\begin{figure}[t]
\begin{center}
\includegraphics[width=0.97\linewidth]{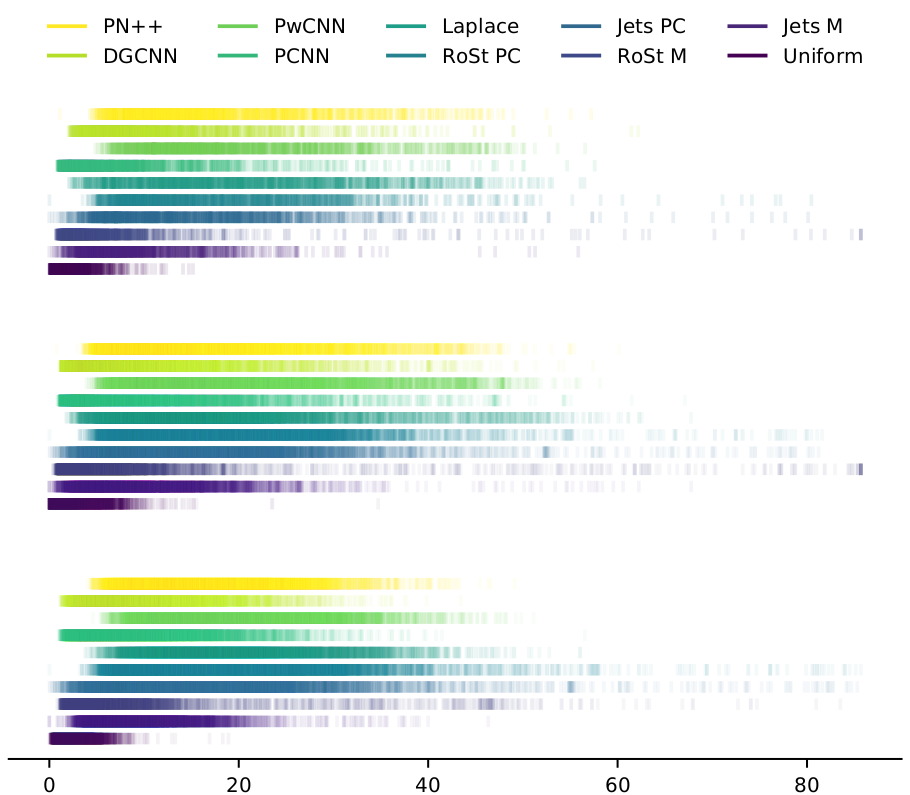}
\end{center}
\vspace{-3mm}
   \caption{Plot of angle deviation error for the high-resolution (2048 points) full model benchmark, using different sample size (top to bottom: 10k, 50k, and 100k).\vspace{-3mm}}
\label{fig:angledev_patch}
\end{figure}

We observe, as expected, that the error is reduced as we increase the number of samples in the training set, this is consistent for all methods on both patches and full models. However, the improvement is modest (Figures ~\ref{fig:angledev_data} and ~\ref{fig:angledev_patch}).

\paragraph{Sampling Resolution on Full Models.}
We also explore how data-driven methods behave when sampling resolution is growing. DGCNN, PCNN, and PwCNN clearly benefit from sampling resolution, while PN++ does not show clear improvements. This phenomenon is likely linked to the spatial subsampling mechanism that is used to ensure sublinear time in training, but prevents this method from leveraging the extra resolution. In case of Laplace surface network, it is difficult to understand the effect since it did not converge after 3 days of training on the highest resolution.

\paragraph{Comparison of Analytic Methods.}
Analytic methods are remarkably consistent across dataset sizes and improve as the mesh sampling resolution is increased. The methods based on surface connectivity heavily outperform those relying on K-nearest neighbour estimation, demonstrating that connectivity is a valuable information for this task.

\paragraph{Data-Driven vs. Analytic Methods.}
Almost all data-driven methods perform well against analytic methods for point clouds, especially if the model resolution is low. However, if the analytic methods are allowed to use connectivity information, they outperform all learning methods by a large margin, even those also using connectivity information. To further support this conclusion, we run a similar experiment on a simpler, synthetic dataset composed of 10k and 50k random NURBS patches and observe similar results, which are available in the supplementary material.

\begin{figure}[t]
\begin{center}
\includegraphics[width=0.97\linewidth]{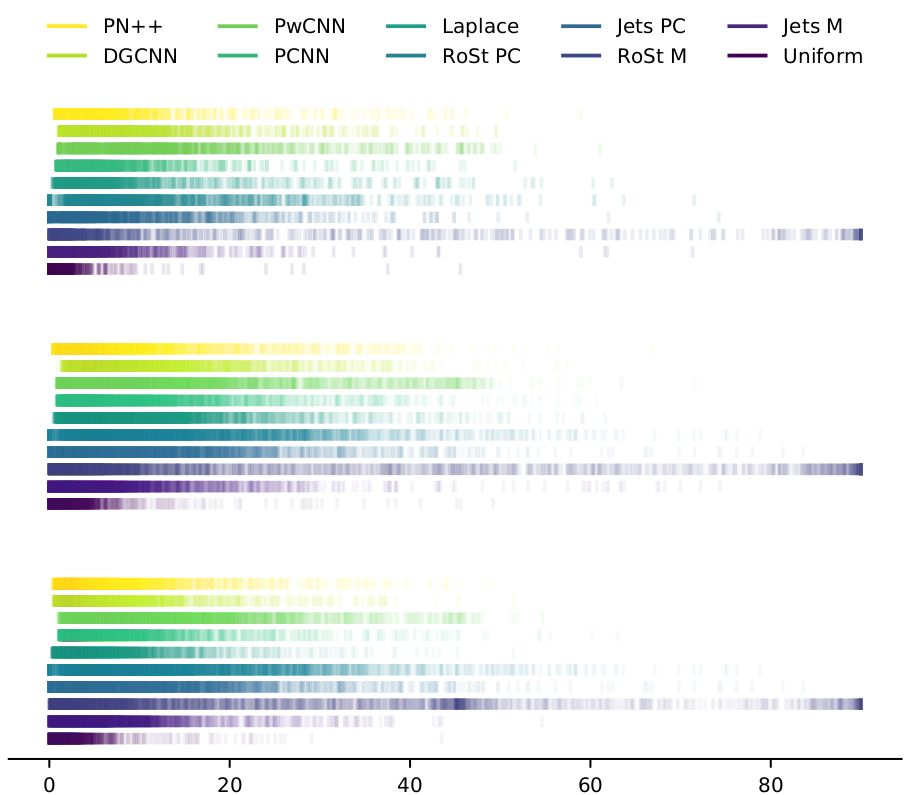}
\end{center}
\vspace{-3mm}
   \caption{Plot of angle deviation error for the lower resolution patches (512 points) benchmark, using different sample size (top to bottom: 10k, 50k, and 100k).}
\label{fig:angledev_data}
\end{figure}

\begin{table*}
\scalebox{0.9}{
\begin{minipage}[]{1.0\linewidth}
\centering
\begin{tabular}{rrrrrrrrrrrrrrrrrr}
\multicolumn{2}{l}{Method/}  && \multicolumn{7}{c}{Full Models} & \phantom{a}& \multicolumn{7}{c}{Patches}\\
\multicolumn{2}{l}{Vertices} && \multicolumn{3}{c}{Loss} && \multicolumn{3}{c}{Angle Deviation [$^\circ$]}
 && \multicolumn{3}{c}{Loss} && \multicolumn{3}{c}{Angle Deviation [$^\circ$]}\\
\cmidrule{4-6}
\cmidrule{8-10} 
\cmidrule{12-14}
\cmidrule{16-18}
&&& 10k & 50k & 100k && 10k & 50k & 100k
&& 10k & 50k & 100k && 10k & 50k & 100k\\
\midrule
\multirow{3}{*}{\rotatebox[origin=c]{90}{PN++}}
& 512 && 0.168 & 0.155 & 0.142 && 7.83 & 7.32 & 6.43 && 0.034 & 0.032 & 0.025 && 1.70 & 1.42 & 1.10\\
& 1024 && 0.180 & 0.163 & 0.160 && 7.49 & 6.06 & 6.17 && 0.056 & 0.052 & 0.048 && 2.11 & 1.57 & 1.51\\
& 2048 && 0.171 & 0.156 & 0.149 && 6.75 & 5.47 & 5.08 && 0.081 & 0.071 & 0.063 && 2.48 & 1.77 & 1.44\\
\midrule
\multirow{3}{*}{\rotatebox[origin=c]{90}{DGCNN}}
& 512 && 0.177 & 0.167 & 0.144 && 9.61 & 8.32 & 7.13 && 0.054 & 0.049 & 0.025 && 3.20 & 3.00 & 1.12\\
& 1024 && 0.126 & 0.104 & 0.099 && 5.91 & 4.59 & 4.34 && 0.048 & 0.036 & 0.024 && 2.98 & 2.15 & 1.13\\
& 2048 && 0.090 & 0.070 & 0.068 && 4.54 & 2.80 & 2.77 && 0.045 & 0.035 & 0.023 && 2.66 & 1.95 & 0.98\\
\midrule
\multirow{3}{*}{\rotatebox[origin=c]{90}{PwCNN}}
& 512 && 0.273 & 0.260 & 0.252 && 18.73 & 17.27 & 16.36 && 0.092 & 0.069 & 0.067 && 4.71 & 3.45 & 3.43\\
& 1024 && 0.217 & 0.218 & 0.198 && 12.78 & 13.10 & 11.38 && 0.107 & 0.110 & 0.089 && 5.50 & 6.11 & 4.58\\
& 2048 && 0.188 & \textit{0.176*} &\textit{0.168*}&& 11.34 & \textit{10.54*} & \textit{10.05*} && 0.107 & \textit{0.120*}& 0.094 && 5.98 & \textit{6.36*} & 4.83\\
\midrule
\multirow{3}{*}{\rotatebox[origin=c]{90}{PCNN}}
& 512 && 0.146 & 0.153 & 0.139 && 6.47 & 6.96 & 6.15 && 0.037 & 0.043 & 0.028 && 1.84 & 1.84 & 1.42\\
& 1024 && 0.104 & 0.099 & 0.103 && 3.56 & 3.46 & 3.69 && 0.025 & 0.030 & 0.025 && 0.94 & 1.37 & 0.92\\
& 2048 && 0.065 & 0.070 & 0.067 && 2.05 & 2.44 & 2.22 && 0.023 & 0.025 & \textit{0.023*} && 0.88 & 1.01 & \textit{0.83*}\\
\midrule
\multirow{3}{*}{\rotatebox[origin=c]{90}{Laplace}}
& 512 && 0.282 & 0.203 & 0.133 && 20.01 & 11.94 & 8.47 && 0.041 & 0.047 & 0.022 && 1.93 & 3.13 & 1.12\\
& 1024 && 0.211 & 0.138 &\textit{0.146*}&& 34.24 & 9.43 & \textit{9.85*} && 0.030 & 0.027 & \textit{0.029*} && 1.65 & 1.36 &\textit{ 1.46*}\\
& 2048 && 0.197 & \textit{0.148*} & \textit{0.158*} && 9.99 & \textit{9.95*} & \textit{10.57*} && 0.031 & 0.040 &\textit{ 0.040*} && 1.60 & 1.67 & \textit{1.81*}\\
\midrule
\midrule
\multirow{3}{*}{\rotatebox[origin=c]{90}{PCPNet}}
& 512 && -- & -- & -- && -- & -- & --  && 0.098$\dagger$ & 0.081$\dagger$ & -- && 9.95$\dagger$ & 9.28$\dagger$ & --\\
& 1024 && -- & -- & -- && -- & -- & -- && 0.123$\dagger$ & 0.097$\dagger$ & -- && 13.89$\dagger$ & 9.55$\dagger$ & --\\
& 2048 && -- & -- & -- && -- & -- & -- && 0.142$\dagger$ & 0.200$\dagger$ & -- && 16.24$\dagger$ & 16.45$\dagger$ & --\\
\midrule
\multirow{3}{*}{\rotatebox[origin=c]{90}{ExtOp}}
& 512 && -- & -- & -- && -- & -- & --  && 0.074$\dagger$ & 0.073$\dagger$ & -- && 2.42$\dagger$ & 2.05$\dagger$ & --\\
& 1024 && -- & -- & -- && -- & -- & -- && 0.095$\dagger$ & 0.096$\dagger$ & -- && 3.32$\dagger$ & 2.50$\dagger$ & --\\
& 2048 && -- & -- & -- && -- & -- & -- && 0.091$\dagger$ & --    & -- && 3.00$\dagger$ & --  & --\\
\midrule
\midrule
\multirow{3}{*}{\rotatebox[origin=c]{90}{RoSt PC}}
& 512 && 0.298 & 0.300 & -- && 21.32 & 21.36 & -- && 0.083 & 0.082 & -- && 0.82 & 0.79 & --\\
& 1024 && 0.220 & 0.223 & -- && 14.47 & 14.63 & -- && 0.078 & 0.077 & -- && 0.74 & 0.72 & --\\
& 2048 && 0.164 & 0.166 & -- && 9.96 & 10.18 & -- && 0.073 & 0.072 & -- && 0.59 & 0.62 & --\\
\midrule
\multirow{3}{*}{\rotatebox[origin=c]{90}{Jets PC}}
& 512 && 0.260 & 0.261 & -- && 17.84 & 17.97 & -- && 0.050 & 0.050 & -- && 0.05 & 0.05 & --\\
& 1024 && 0.183 & 0.186 & -- && 12.19 & 12.39 & -- && 0.048 & 0.048 & -- && 0.05 & 0.05 & --\\
& 2048 && 0.129 & 0.132 & -- && 8.41 & 8.63 & -- && 0.045 & 0.044 & -- && 0.04 & 0.04 & --\\
\midrule
\multirow{3}{*}{\rotatebox[origin=c]{90}{RoSt M}}
& 512 && 0.082 & 0.084 & 0.084 && 2.15 & 2.17 & 2.18 && 0.108 & 0.103 & 0.102 && 0.06 & 0.06 & 0.06\\
& 1024 && 0.053 & 0.055 & 0.056 && 0.25 & 0.29 & 0.29 && 0.107 & 0.105 & 0.105 && 0.06 & 0.06 & 0.06\\
& 2048 && 0.047 & 0.048 & 0.050 && 0.08 & 0.08 & 0.08 && 0.112 & 0.108 & 0.107 && 0.06 & 0.06 & 0.06\\
\midrule
\multirow{3}{*}{\rotatebox[origin=c]{90}{Jets M}}
& 512 && 0.175 & 0.176 & 0.175 && 7.26 & 7.29 & 7.29 && 0.036 & 0.036 & 0.036 && 0.00 & 0.00 & 0.00\\
& 1024 && 0.118 & 0.118 & 0.117 && 0.10 & 0.10 & 0.11 && 0.033 & 0.033 & 0.033 && 0.00 & 0.00 & 0.00\\
& 2048 && 0.078 & 0.079 & 0.079 && 0.01 & 0.02 & 0.02 && 0.029 & 0.031 & 0.031 && 0.00 & 0.00 & 0.00\\
\midrule
\multirow{3}{*}{\rotatebox[origin=c]{90}{Uniform}}
& 512 && 0.024 & 0.025 & 0.024 && 0.26 & 0.29 & 0.28 && 0.007 & 0.007 & 0.007 && 0.00 & 0.00 & 0.00\\
& 1024 && 0.013 & 0.013 & 0.013 && 0.00 & 0.00 & 0.00 && 0.005 & 0.005 & 0.005 && 0.00 & 0.00 & 0.00\\
& 2048 && 0.009 & 0.010 & 0.009 && 0.00 & 0.00 & 0.00 && 0.004 & 0.005 & 0.004 && 0.00 & 0.00 & 0.00\\
\midrule

\end{tabular}
\end{minipage}}
\caption{Statistical results for all evaluated methods for the full model and patch benchmarks.
The loss is calculated as $1 - \left(\mathbf{n}^\mathsf{T} \mathbf{\hat{n}}\right)^2$ and the angle deviation is calculated as the angle in degrees $\angle(\mathbf{n}, \mathbf{\hat{n}})$ between ground truth normal and estimated normal. 
For the loss we report the mean over all models, for the angle deviation we report the median of all models in the according datasets.
Osculating Jets and Robust Statistical Estimation are evaluated both on point cloud inputs (PC suffix; comparable to the learning methods) as well as mesh inputs (M suffix). 
$\dagger$: PCPNet and ExtOp were not run on full models since they compute only 1 normal per patch (and their loss, differently from all other rows, is computed only on the vertex in the center of the patch).
*: the training was not completed before the time limit is reached, and the partial result is used for inference.}
\label{tab:results}
\end{table*}

\section{Conclusion}

We introduced a new large dataset of CAD models, and a set of tools to convert them into the representations used by deep learning algorithms. The dataset will continue to grow as more models are added to the Onshape public collection. Large scale learning benchmarks can be created using the ground truth signals that we extract from the CAD data, as we demonstrate for the estimation of differential surface quantities. 

The result of our comparison will be of guidance to the development of new geometric deep learning methods. Our surprising conclusion is that, while deep learning methods which use only 3D points are superior to analytical methods, this is not the case when connectivity information is available. This suggests that existing graph architectures struggle at exploiting the connectivity efficiently and are considerably worse than the simplest analytical method (uniform), which simply averages the normals of neighbouring faces. It would be interesting to run a similar study by extending these algorithms to correctly identify and predict multiple normals on sharp features and compare them with specialized methods for this task \cite{boulch_cnn}.

Another surprising discovery is that even the uniform algorithm fails to produce valid normals on roughly 100 models in our dataset due to floating point errors. These kinds of problems are extremely challenging to identify, and we believe that the size and complexity of our dataset are an ideal stress test for robust geometry processing algorithms.

\section{Distribution}
\label{ssec:distribution}

The dataset and all information is available at:\\ {\small\url{https://deep-geometry.github.io/abc-dataset}}
It is distributed under the MIT license and split into chunks of 10k models for each data type.

The copyright owners of the models are the respective creators (listed in the meta information).
The geometry processing pipeline is made available under the GPL license in form of a containerized software solution (Docker \cite{docker} and Singularity \cite{singularity}) that can be run on every suitable machine.

\section*{Acknowledgements}
{
\small
We are grateful to Onshape for providing the CAD models and support. This work was supported in part through the NYU IT High Performance Computing resources, services, and staff expertise. Funding provided by NSF award MRI-1229185. We thank the Skoltech CDISE HPC Zhores cluster staff for computing cluster provision. This work was supported in part by NSF CAREER award 1652515, the NSF grants IIS-1320635, DMS-1436591, and 1835712, the Russian Science Foundation under Grant 19-41-04109, and gifts from Adobe Research, nTopology Inc, and NVIDIA.
}

{\small
\bibliographystyle{ieee}
\bibliography{main}
}

\newpage

\begin{table*}
\begin{minipage}[]{1.0\linewidth}
\centering
\Large
\textbf{ABC: A Big CAD Model Dataset For Geometric Deep Learning}\\\textbf{Supplementary Material}
\end{minipage}
\end{table*}

\begin{appendices}

\begin{figure}[h]
\begin{center}
\includegraphics[width=1.0\linewidth]{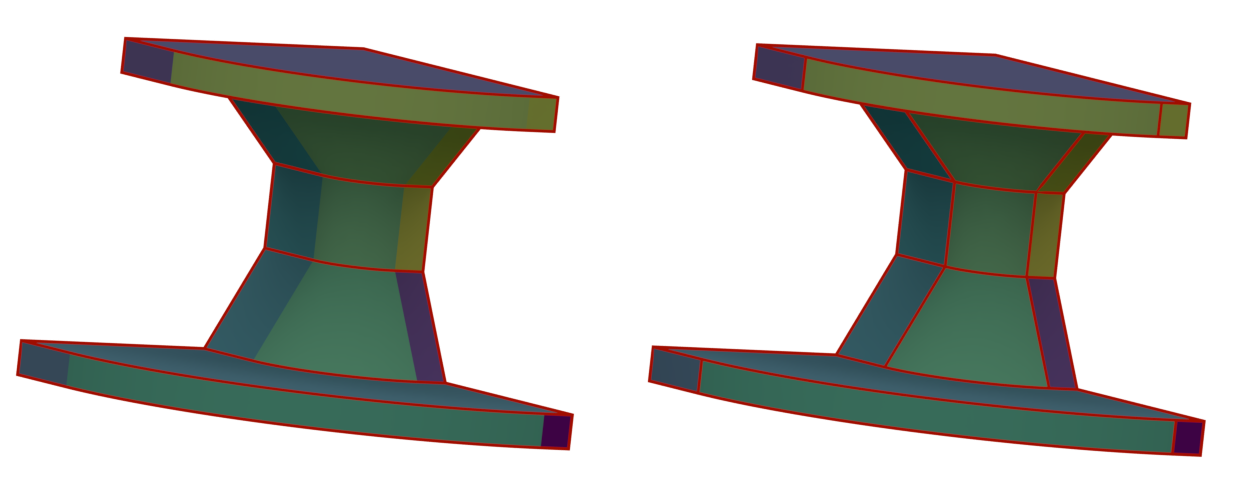}
\end{center}
   \caption{Example model with differently colored patches and highlighted sharp feature curves on the left as well as all feature curves on the right.}
\label{fig:typedistribution}
\end{figure}

\section{Model Filtering and Post-Processing}
We filter out defective and low quality models in the Onshape collection using a set of automatic filters, which users can modify depending on their application. First, empty CAD models are filtered by file size of the original STEP files. Second, models that only consist of a single primitive or models that require shape healing during the translation and meshing phase are also filtered out. At this stage, we also filter by file size of the resulting meshes to avoid extremely large meshes that occur in some corner cases, where the meshing algorithm overly refines the CAD model. 
For the benchmark datasets we have two additional post-processing steps. The full models are generated by resampling the triangle meshes to match the defined numbers of vertices \cite{Garland:1997}. For the patches, we perform breadth-first search starting from a random selection of vertices to generate patches of pre-defined sizes.

\section{File Types Description}
Every model in the ABC-Dataset is stored in three different representations and multiple filetypes.

\subsection{Boundary Representation/CAD}
This is the original format acquired from Onshape, which contains an explicit description of the topology and geometry information of the CAD models in STEP and Parasolid format. The STEP files can be read and processed with Open Cascade \cite{Opencascade} and Gmsh \cite{gmsh}. The processing allows for example to sample at arbitrary resolution, to generate meshes and to extract differential quantities.

\subsection{Discrete Triangle Meshes}
The discrete triangle meshes are supplied in two formats. The first is an STL file which is generated from the Parasolid format by Onshape with a high resolution. While these meshes are faithful approximations of the original geometry, the mesh quality is low: the triangles may have bad aspect ratios, and the sampling can be highly non-uniform, which is undesirable for many geometry processing algorithms. We thus also provide a second mesh, in OBJ format, produced by our processing pipeline. Our result is fairly regular, with a uniform vertex distribution and most triangles have angles close to $60^\circ$. In addition to the triangle mesh itself, differential properties are analytically derived from the boundary representation and stored in these OBJ files. The vertices and faces of the OBJ are matched with the curves and patches stored in the YAML representation described in Section \ref{sec:yaml}. Note that OBJ uses 1-indexing of the vertices, whereas we use 0-indexing in YAML.

\subsection{Curve and Patch Features}
\label{sec:yaml}
The boundary representation of the STEP files defines surfaces and curves of different types. 
In addition to the geometrical information in the files listed above, we store the defining properties of surfaces and curves with references to the corresponding vertices and faces of the discrete triangle mesh representation.
All this information is stored in YAML files \cite{yaml}, which contain a list of patches, and a list of curves, describing the boundary of the patches.
Figure~\ref{fig:typedistribution} shows one example model where different patches are highlighted in different colors and feature curves are drawn as red lines, all loaded from the OBJ and YAML files.

\paragraph{Curves.}
The curves list contains all curves of the CAD model.
For each curve, different information is given depending on its type.

\begin{description}
\item[type] Line, Circle, Ellipse, BSpline, Other.
\item[sharp] True if this curve is a sharp feature curve.
\item[vert\_indices] List of all mesh vertex indices that are sampled from the curve (0-indexed).

\item[vert\_parameters] List of the parameters that describe the corresponding mesh vertices.
\item[Line]{ 
    $\mv{c}(t) = \mv{l} + t \cdot \mv{d}$
    \begin{itemize}
        \item location($\mv{l}$): The location vector of the line.
        \item direction($\mv{d}$): The direction vector of the line.
    \end{itemize}
}
\item[Circle]{
    $\mv{c}(t) = \mv{l} + r \cdot \cos(t) \cdot \mv{x} + r \cdot \sin(t) \cdot \mv{y}$
    \begin{itemize}
        \item location($\mv{l}$): The center of the circle.
        \item z\_axis: The normal axis of the plane of the circle.
        \item radius($r$): The radius of the circle.
        \item x\_axis($\mv{x}$): The first axis of the local coordinate system.
        \item y\_axis($\mv{y}$): The second axis of the local coordinate system.
    \end{itemize}
}
\item[Ellipse]{
    $\mv{c}(t) = \mv{l} + r_x \cdot \cos(t) \mv{x} + r_y \cdot \sin(t) \cdot \mv{y}$
    \begin{itemize}
        \item focus1: The first focal point of the ellipse.
        \item focus2: The second focal point of the ellipse. 
        \item x\_axis($\mv{x}$): The longer/major axis of the ellipse.
        \item y\_axis($\mv{y}$): The shorter/minor axis of the ellipse.
        \item z\_axis: The normal axis of the plane of the ellipse.
        \item x\_radius($r_x$): The major radius of the ellipse.
        \item y\_radius($r_y$): The minor radius of the ellipse.
    \end{itemize}
}
\item[BSpline]{
    Spline curves defined by control points, knots, and optionally weights
    \begin{itemize}
        \item rational: True if the B-Spline is rational.
        \item closed: True if the B-Spline describes a closed curve.
        \item continuity: The order of continuity of the B-Spline functions.
        \item degree: The degree of the B-Spline polynomial functions.
        \item poles: The control points of the B-Spline.
        \item knots: The knot vector with duplicate knots in case of multiplicity greater than 1.
        \item weights: The weights of the B-Spline curve (only used if it is a rational NURBS curve).
    \end{itemize}
}
\end{description}

\paragraph{Patches.}
The patches list contains all patches of the CAD model.
For each patch, different information is given depending on its type.

\begin{description}
\item[type] Plane, Cylinder, Cone, Sphere, Torus, Revolution, Extrusion, BSpline, Other.
\item[vert\_indices] List of all mesh vertex indices that are part of the patch (0-indexed).
\item[vert\_parameters] List of the parameters that describe the according mesh vertices.
\item[face\_indices] List of all face indices that are part of the patch (0-indexed).
\item[Plane]{ 
    $\mv{p}(u, v) = \mv{l} + u \cdot \mv{x} + v \cdot \mv{y}$
    \begin{itemize}
        \item location($\mv{l}$): The location vector of the plane.
        \item x\_axis($\mv{x}$): The first axis of the plane coordinate system.
        \item y\_axis($\mv{y}$): The second axis of the plane coordinate system.
        \item z\_axis: The normal axis of the plane.
        \item coefficients: Coefficients for the cartesian description of the plane: $c[0] \cdot x + c[1] \cdot y + c[2] \cdot z + c[3] = 0.0$. 
    \end{itemize}
}
\item[Cylinder]{ 
    $\mv{p}(u, v) = \mv{l} + r \cdot cos(u) \cdot \mv{x} + r \cdot sin(u) \cdot \mv{y} + v \cdot \mv{z}$
    \begin{itemize}
        \item location($\mv{l}$): The location vector defining the base plane.
        \item x\_axis($\mv{x}$): The first axis of the cylinder coordinate system.
        \item y\_axis($\mv{y}$): The second axis of the cylinder coordinate system.
        \item z\_axis($\mv{z}$): The rotation/center axis of the cylinder.
        \item coefficients: Coefficients for the cartesian quadric description of the cylinder: $c[0] \cdot x^2 + c[1] \cdot y^2 + c[2] \cdot z^2 + 2\cdot (c[3] \cdot x \cdot y + c[4] \cdot x \cdot z + c[5] \cdot y \cdot z) + 2\cdot (c[6] \cdot x + c[7] \cdot y + c[8] \cdot z) + c[9] = 0.0$. 
    \end{itemize}
}
\item[Cone]{ 
    $\mv{p}(u, v) = \mv{l} + (r + v \cdot \sin(a)) \cdot (\cos(u) \cdot \mv{x} + \sin(u) \cdot \mv{y}) + v \cdot \cos(a) \cdot \mv{z}$
    \begin{itemize}
        \item location($\mv{l}$): The location vector defining the base plane.
        \item x\_axis($\mv{x}$): The first axis of the cone coordinate system.
        \item y\_axis($\mv{y}$): The second axis of the cone coordinate system.
        \item z\_axis(z): The rotation/center axis of the cone.
        \item coefficients: Coefficients for the Cartesian quadric description of the cone: $c[0] \cdot x^2 + c[1] \cdot y^2 + c[2] \cdot z^2 + 2\cdot (c[3] \cdot x \cdot y + c[4] \cdot x \cdot z + c[5] \cdot y \cdot z) + 2\cdot (c[6] \cdot x + c[7] \cdot y + c[8] \cdot z) + c[9] = 0.0$. 
        \item radius($r$): The radius of the circle that describes the intersection of the cone and base plane.
        \item angle($a$): The half-angle at the apex of the cone.
        \item apex: The apex/tip of the cone.
    \end{itemize}
}
\item[Sphere]{ 
    $\mv{p}(u, v) = \mv{l} + r \cdot cos(v) \cdot (cos(u) \cdot \mv{x} + sin(u) \cdot \mv{y}) + r \cdot sin(v) \cdot \mv{z}$
    \begin{itemize}
        \item location($\mv{l}$): The location vector defining center of the sphere.
        \item x\_axis($\mv{x}$): The first axis of the sphere coordinate system.
        \item y\_axis($\mv{y}$): The second axis of the sphere coordinate system.
        \item z\_axis($\mv{z}$): The third axis of the sphere coordinate system.
        \item coefficients: Coefficients for the Cartesian quadric description of the sphere: $c[0] \cdot x^2 + c[1] \cdot y^2 + c[2] \cdot z^2 + 2\cdot (c[3] \cdot x \cdot y + c[4] \cdot x \cdot z + c[5] \cdot y \cdot z) + 2\cdot (c[6] \cdot x + c[7] \cdot y + c[8] \cdot z) + c[9] = 0.0.$ 
        \item radius($r$): The radius of the sphere.
    \end{itemize}
}
\item[Torus]{ 
    $\mv{p}(u, v) = \mv{l} + (r_{max} + r_{min} \cdot cos(v)) \cdot (cos(u) \cdot \mv{x} +\sin(u) \cdot \mv{y}) + r \cdot \sin(v) \cdot \mv{z}$
    \begin{itemize}
        \item location($\mv{l}$): The location  defining center of the torus.
        \item x\_axis($\mv{x}$): The first axis of the torus coordinate system.
        \item y\_axis($\mv{y}$): The second axis of the torus coordinate system.
        \item z\_axis($\mv{z}$): The rotation/center axis of the torus.
        \item max\_radius($r_{max}$): The major/larger radius of the torus.
        \item min\_radius($r_{min}$): The minor/smaller radius of the torus.
    \end{itemize}
}
\item[Revolution]{ 
    Surface of revolution: a curve is rotated around the rotation axis.
    \begin{itemize}
        \item location: A point on the rotation axis
        \item z\_axis: The rotation axis dirction.
        \item curve: The rotated curve that can be of any of the curve types.
    \end{itemize}
}
\item[Extrusion]{ 
    Surface of linear extrusion: a curve is extruded along a direction. 
    \begin{itemize}
        \item direction: The linear extrusion direction of the surface ($v$ parameter).
        \item curve: The extruded curve that can be of any of the curve types ($u$ parameter).
    \end{itemize}
}
\item[BSpline]{
    Spline patch defined by control points, knots, and optionally weights.
    \begin{itemize}
        \item u\_rational: True if the B-Spline is rational in $u$ direction.
        \item v\_rational: True if the B-Spline is rational in $v$ direction.
        \item u\_closed: True if the B-Spline describes a closed surface in $u$ direction.
        \item v\_closed: True if the B-Spline describes a closed surface in $v$ direction.
        \item continuity: The order of continuity of the B-Spline functions.
        \item u\_degree: The degree of the B-Spline polynomial functions in $u$ direction.
        \item v\_degree: The degree of the B-Spline polynomial functions in $v$ direction.
        \item poles: 2D array of control points. The first dimension corresponds to the $u$ direction, the second dimension to the $v$ direction.
        \item u\_knots: The knot vector for $u$ with duplicate knots in case of multiplicity greater than 1.
        \item v\_knots: The knot vector for $v$ with duplicate knots in case of multiplicity greater than 1.
        \item weights: 2D array of the weights of the NURBS patch, corresponding to the control points (only used if the patch is rational).
    \end{itemize}
}
\end{description}

\section{Implementation Details}

\begin{table*}
\begin{center}
\begin{tabular}{lccccccc}
Method & PN++ & DGCNN & PwCNN & PCNN & Laplace & PCPN & ExtOp\\
\midrule
Parameters & 1402180 & 1724100 & 10207 & 11454787 & 1003395 & 3469255 & 8189635 \\
\end{tabular}
\end{center}
\vspace{-3mm}
\caption{Overview of the network capacities for the machine learning approaches.}
\label{tab:capacity}
\end{table*}

\paragraph{Parameters.}
The overall number of parameters used by each method is listed in Table~\ref{tab:capacity}, while the running time is listed in Table \ref{tab:timings} in this document. In the following we give a brief overview of the modifications and settings we used for each method in our comparison. One common change that we performed on all methods is to switch to the cosine loss described in the main article (Section 5.1), and we provided a maximum allowed time of 3 days. 

\paragraph{Point Convolutional Neural Networks by Extension Operators \cite{pcnnExt}.}
The architecture is the same as the one proposed for classification, but with the last layer producing 3 values instead of 10. For each of the input sizes, we changed the size of the input layer accordingly. 
We used a step size of $10^{-3}$ using stochastic gradient descent. For the datasets with patch sizes of 512 and 1024, we used a minibatch size of 32. For the datasets with a patch size of 2048, we used a minibatch size of 16. We trained each network for up to 250 epochs.

\paragraph{Surface Networks \cite{surface_networks}.}
In some rare cases, there are degenerate triangles in the triangle meshes, which pose challenge for discrete Laplacian operator computation. When the coefficients overflow, we replace these coefficients with 1. For training, we use 300 epochs, with Adam \cite{kingma2014adam} optimizer and learning rate starting from $10^{-3}$, and after 100 epochs, halved at every 20 epochs.

\paragraph{PointNet++: Deep Hierarchical Feature Learning on Point Sets in a Metric Space \cite{pointnetpp}.} Since no experiments for normal estimation were provided in the paper, we decided that the most natural way to predict normals with PointNet is by modifying the segmentation architecture, predicting three continuous numbers and normalizing them. We trained for 100 epochs with default settings and batch size 16.

\paragraph{Dynamic Graph CNN for Learning on Point Clouds \cite{dgcnn}.} We adapted the segmentation architecture by changing the last layer of the model to output 3 continuous values. Normalization was applied to ensure the output vector has unit length. We trained for 100 epochs with the default settings used for segmentation in the original implementation, batch size 16.

\paragraph{PCPNet: Learning Local Shape Properties from Raw Point Clouds \cite{pcpnet}.} We used exactly the same architecture specified in the original paper \cite{pcpnet}. We trained with the Adam \cite{kingma2014adam} optimizer using a step size of $1 \times 10^{-3}$, $\beta = (0.9, 0.99)$, and $\epsilon = 1 \times 10^{-8}$, and no weight decay. Training was run for up to 2000 epochs.

\paragraph{Pointwise Convolutional Neural Networks \cite{pwcnn}.} We adapted PwCNN in the same fashion as PointNet++ and DGCNN. We used 100 epochs for training, and took default segmentation settings from the original implementation with batch size 16.

\paragraph{PointCNN \cite{pointcnn}.}
We took the PointCNN segmentation architecture, changed the output dimensionality to 3, and ran the training procedure with 100 epochs and default settings, batch size 8.

\paragraph{Osculating Jets \cite{osculatingjets}.} 
For the mesh version we use the default parameters of the CGAL implementation. In the point cloud version, we use the 10 nearest neighbours for the jet fitting.

\paragraph{Robust Statistical Estimation on Discrete Surfaces \cite{discrete_surf}.}
For the mesh version of RoSt we use the default parameters set by the authors. The point cloud version is supplied with estimated normals from locally fitting a plane to the 10 nearest neighbours of each point. The method then refines these normals.

\section{Running Times}
We report the training time for data-driven methods in Table \ref{tab:training}, and the running times for analytic methods in Table \ref{tab:timings}. We capped the \textit{training} time to three days, although the instances are trained on different machines. The colors in the table denote the GPU model that has been used for the training. For the analytic methods, the times are measured for running the normal estimation on one CPU core (Intel Core i7).









\begin{table}
\scalebox{0.85}{
\begin{minipage}[]{1.0\linewidth}
\centering
\begin{tabular}{rrrrrrrrrr}
\multicolumn{2}{l}{Method / \#V} && \multicolumn{3}{c}{Full Models} 
 && \multicolumn{3}{c}{Patches}\\
\cmidrule{4-6}
\cmidrule{8-10} 
&&& 10k & 50k & 100k && 10k & 50k & 100k\\
\midrule
\multirow{3}{*}{\rotatebox[origin=c]{90}{PN++}}
& 512 && 173& 857 & \Vcent{1319} && 173&871&\Vcent{1241}  \\
& 1024 &&184	&{919}	&\Vcent{1615}&&183	&920	&\Vcent{1613}  \\
& 2048 && 226&	{1196}&	\Vcent{2124}&&227&	1139&	\Vcent{2328}\\
\midrule
\multirow{3}{*}{\rotatebox[origin=c]{90}{DGCNN}}
& 512 && 155&	734	&\Vcent{819}&&	155&	784&	\Vcent{814} \\
& 1024 &&299&	1158&	\Vcent{1257}&&	222&	1163&	\Vcent{1268}\\
& 2048 &&490&	2462&	\Vcent{2467}&&	487&	2401&	\Vcent{2498} \\
\midrule
\multirow{3}{*}{\rotatebox[origin=c]{90}{PwCNN}}
& 512  &&258	&1173&	\Vcent{971}	&&244&	1256	&\Vcent{891} \\
& 1024 &&515 &	2626&	\Vcent{2027}&&	471&	3413&	\Vcent{1773}\\
& 2048 &&1293&	TO&	\Vcent{TO}&&	1319&	TO&	\Vcent{4179} \\
\midrule
\multirow{3}{*}{\rotatebox[origin=c]{90}{PCNN}}
& 512 &&614&	3037&	\Vcent{3375}	&&616&	3071&	\Vcent{3483} \\
& 1024 &&675&	3370&	\Vcent{3599}&&	659&	3368&	\Vcent{3727}\\
& 2048 && 862&	4248&	\Vcent{4298}&&	848	&4110&	\Vcent{TO}\\
\midrule
\multirow{3}{*}{\rotatebox[origin=c]{90}{Laplace}}
& 512 && \TtV{492}	&1438&	\TtXp{2793}&&	285	&\TtXp{1402}	&\TtXp{2747}\\
& 1024 &&\TtX{1001}&	\Mix{31}{34}&	\Pcent{TO}&&	\TtX{983}	&\TtXp{2845}	&\Mix{T}{O}\\
& 2048 && \TtX{1939}&	\Mix{T}{O}&	\Mix{T}{O}&&	1251&	2807&	\TtXp{TO}\\
\midrule
\end{tabular}
\end{minipage}}
\caption{Training time (in minutes) for all evaluated methods for the full model and patch benchmarks.
Coloring indicates different NVidia GPUs, with
default \TenEighty{GTX 1080Ti}, \Vcent{Tesla V100}, \Pcent{Tesla P100}, \TtV{Titan V}, \TtX{Titan X}, \TtXp{Titan Xp}.
TO: training process time-out of 3-day (4320 min) limit. This occurred for all trainings of PCPNet and ExtOp.
*: Using more than two GPU (serially) during single training instance, typically with GTX 1080Ti and \TtXp{Titan Xp}.}
\label{tab:training}
\end{table}

\begin{table}
\scalebox{0.85}{
\begin{minipage}[]{1.0\linewidth}
\centering
\begin{tabular}{rrrrrrrrrr}
\multicolumn{2}{l}{Method / \#V} && \multicolumn{3}{c}{Full Models} 
 && \multicolumn{3}{c}{Patches}\\
\cmidrule{4-6}
\cmidrule{8-10} 
&&& 10k & 50k & 100k && 10k & 50k & 100k\\
\midrule
\multirow{3}{*}{\rotatebox[origin=c]{90}{RoSt PC}}
& 512 && 9.6 & 48.7 & --       && 9.2 & 46.9& --\\
& 1024 && 17.4 & 87.1 & --    && 16.7 & 81.8& --\\
& 2048 && 32.6 & 161.6 & --   && 30.4 & 148.1& --\\
\midrule
\multirow{3}{*}{\rotatebox[origin=c]{90}{Jets PC}}
& 512 && 6.5 & 32.3 & --      && 5.9& 29.7& --\\
& 1024 && 12.3& 61.6 & --     && 11.3& 57.63& --\\
& 2048 && 23.6& 117.8 & --    && 22.1& 114.9&  --\\
\midrule
\multirow{3}{*}{\rotatebox[origin=c]{90}{Uniform}}
& 512 && 0.5 & 2.8 & 5.7 && 0.4& 2.7& 6.6\\ 
& 1024 && 0.8 & 3.9 & 7.9 && 0.6& 2.9& 8.8\\
& 2048 && 1.1 & 6.0 & 12.0 && 1.1& 4.6& 12.1\\
\midrule
\end{tabular}
\end{minipage}}
\caption{Running time (in minutes) for traditional methods for the full model and patch benchmarks.}
\label{tab:timings}
\end{table}

\newpage
\clearpage

\begin{figure}[t]
\begin{center}
\includegraphics[width=0.95\linewidth]{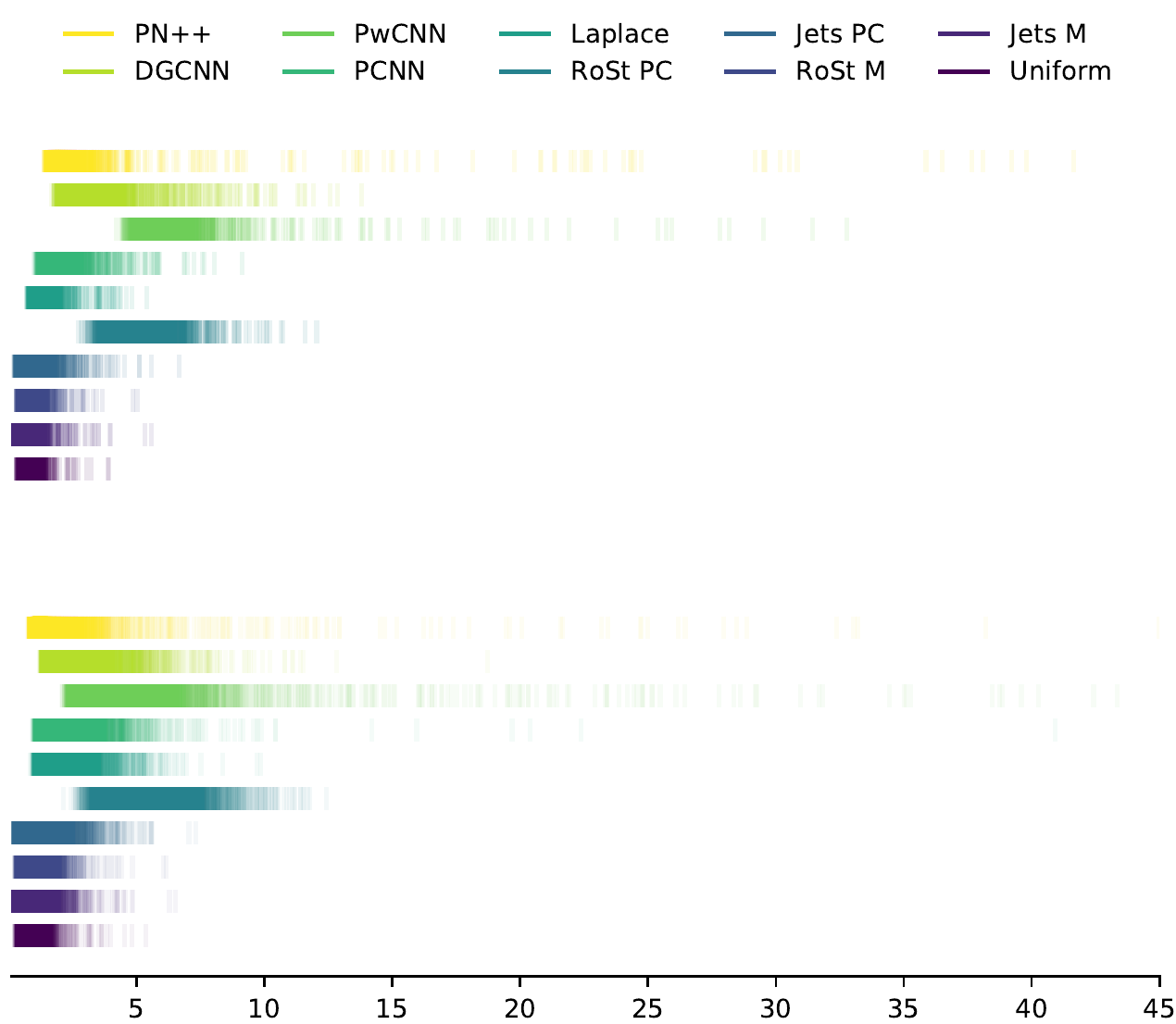}
\end{center}
\vspace{-3mm}
   \caption{Plot of angle deviation error for the lower resolution NURBS (512 points) benchmark, using different sample size (top to bottom: 10k and 50k).}
\label{fig:angledev_data}
\end{figure}

\section{NURBS Patches Experiment}
\begin{figure}[h]
\begin{center}
\includegraphics[width=1.0\linewidth]{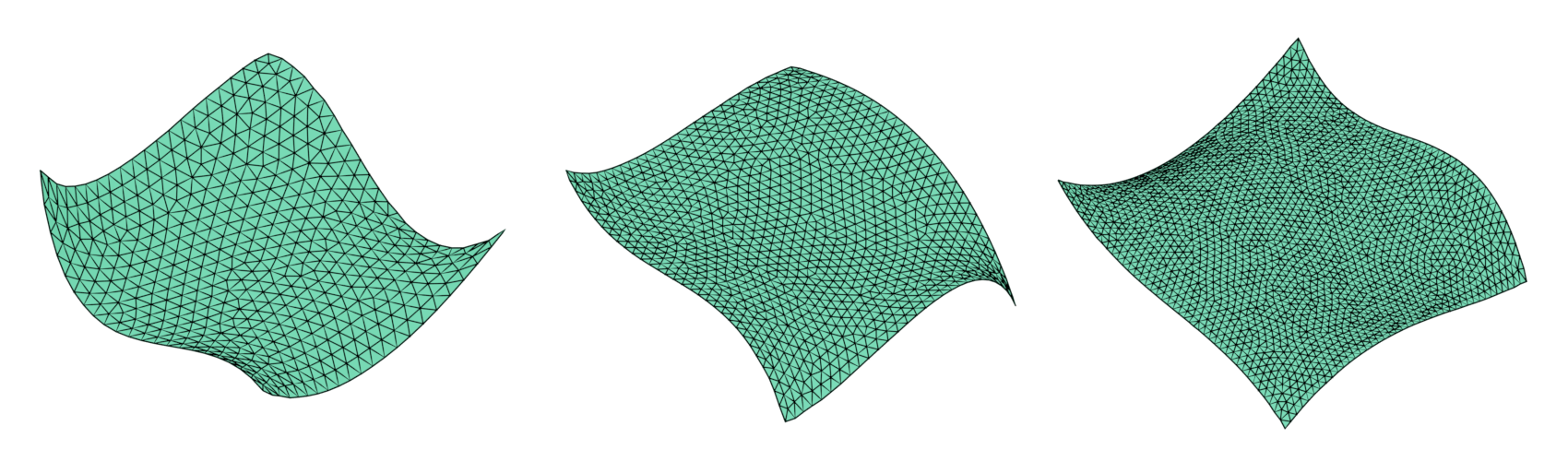}
\end{center}
\vspace{-3mm}
   \caption{Example NURBS patches of the three different sampling densities with 512, 1024, and 2048 vertices (from left to right).}
\label{fig:nurbs}
\end{figure}

We perform an additional experiment on a synthetic dataset, where normals are supposedly easier to infer as the space of possible shapes is artificially restricted, giving an advantage to data-driven methods over traditional ones. We generated two datasets of 10k and 50k random bi-variate B-Spline surface patches of the same three different sampling densities: 512, 1024 and 2048 (see Figure~\ref{fig:nurbs}). Each patch was generated by randomly picking between $0.1\times\text{samples}$ and $2.0\times\text{samples}$ points, choosing random values with uniform probability between 0 and 1, and computing an interpolating bicubic spline surface using the function \texttt{scipy.interpolate.SmoothBivariateSpline} \cite{ScyPySpline}. By construction, the surface is smooth and the normals are thus easier to predict than on real-world geometric models. The relative performance of data-driven and traditional methods is very similar to the one reported in the paper (Section 5), confirming the trend observed on real-world models.

\newpage

\begin{figure}[h]
\begin{center}
\includegraphics[width=0.95\linewidth]{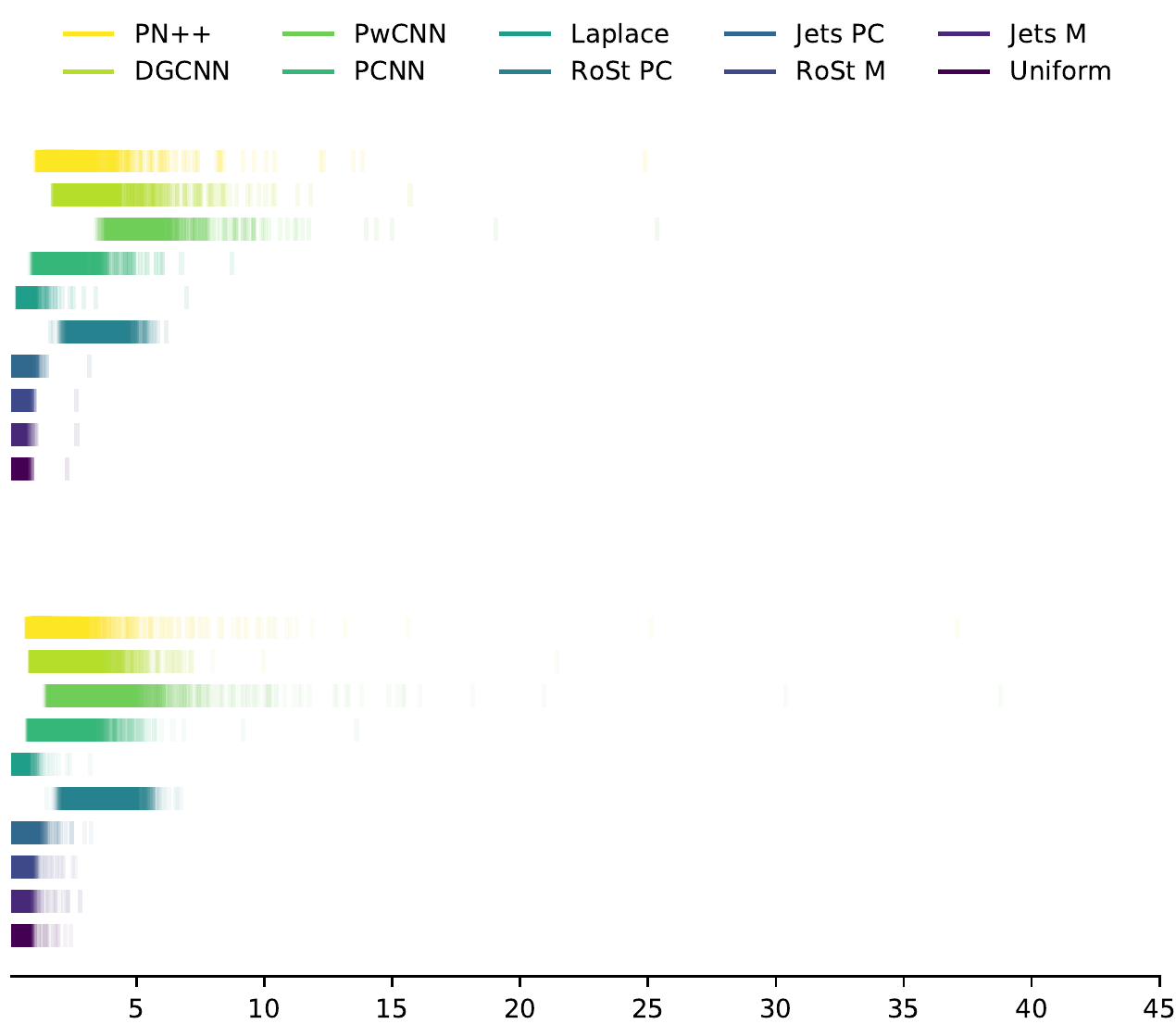}
\end{center}
\vspace{-3mm}
   \caption{Plot of angle deviation error for the medium resolution NURBS (1024 points) benchmark, using different sample size (top to bottom: 10k and 50k).}
\label{fig:angledev_data}
\end{figure}

\end{appendices}

\end{document}